\documentclass[sigconf,nonacm]{acmart}

\usepackage{multirow}
\usepackage{colortbl}  %
\usepackage{xspace}
\usepackage{subcaption}
\usepackage{tikz}
\usetikzlibrary{positioning,fit,backgrounds,calc,shapes.geometric,arrows.meta,decorations.pathreplacing}
\usepackage{pifont}  %
\usepackage{algorithm}
\usepackage[noend]{algpseudocode}

\newcommand{\method}{\textsc{HyperAddress}\xspace}    %
\definecolor{ours}{HTML}{DCE3F0}
\newcommand{\oursrow}{\rowcolor{ours}}
\newcommand{\midruleours}{\specialrule{\lightrulewidth}{\aboverulesep}{0pt}}
\newcommand{\bottomruleours}{\specialrule{\heavyrulewidth}{0pt}{\belowbottomsep}}
\newcommand{\oursrowlast}{\rowcolor{ours}%
  \rule[-\dimexpr\dp\strutbox+\aboverulesep\relax]{0pt}{\dimexpr\ht\strutbox+\dp\strutbox+\aboverulesep+\belowrulesep\relax}}

\begin{document}

\title{Read the Route: Attention-Steered Hypergraph Retrieval for Multi-hop RAG}

\author{Jiate Liu}
\affiliation{%
  \institution{The University of New South Wales}
  \city{Sydney}
  \country{Australia}}
\email{jiate.liu@unsw.edu.au}

\author{Liuyi Chen}
\affiliation{%
  \institution{Hunan University}
  \city{Changsha}
  \country{China}}
\email{liuyi.chen@hnu.edu.cn}

\author{Xiaofei Yuan}
\affiliation{%
  \institution{Euler AI}
  \country{Australia}}
\email{xiaofei.yuan@eulerai.au}

\author{Yifu Tang}
\affiliation{%
  \institution{Vecton AI}
  \country{Australia}}
\email{yves.tang@vectonai.com}

\author{Zihan Yang}
\affiliation{%
  \institution{Vecton AI}
  \country{Australia}}
\email{zihan.yang@vectonai.com}

\author{Xu Zhou}
\affiliation{%
  \institution{Hunan University}
  \city{Changsha}
  \country{China}}
\email{zhxu@hnu.edu.cn}

\author{Zhengyi Yang}
\authornote{Corresponding author.}
\affiliation{%
  \institution{The University of Sydney}
  \city{Sydney}
  \country{Australia}}
\email{zhengyi.yang@sydney.edu.au}

\renewcommand{\shortauthors}{Liu et al.}

\begin{abstract}
Graph-based retrieval-augmented generation (GraphRAG) answers multi-hop questions by organizing a corpus into a graph and retrieving
along its structure. Existing retrieval methods face two limitations. Diffusion- and similarity-based methods expand retrieval along
fixed edge weights or by similarity to the question, without resolving its implicit entity references, allowing relevance errors to
accumulate across hops. In contrast, exploration-based methods let an LLM reason about the next hop, but require a separate
generation at each hop, leading to high inference costs. We observe that an LLM can resolve implicit entity references from
retrieved passages and a multi-hop question before generating any tokens. Further analysis identifies a small set of attention
heads, termed addressing heads, that capture this capability across multiple hops in a single forward pass. Building on this
finding, we propose \method, which runs Personalized PageRank on a description-labelled hypergraph, using addressing heads to
determine the restart distribution and transition weights, thereby guiding multi-hop retrieval without any generation at query time.
On MuSiQue, 2WikiMultiHopQA, and HotpotQA, \method outperforms the strongest baseline by 3.5 points of Recall@5 and 7.4 points of
All@5 on average, with gains that grow with the number of hops. It answers more than twice as fast and builds its index at half the
cost.

\end{abstract}

\begin{CCSXML}
<ccs2012>
   <concept>
       <concept_id>10002951.10003317.10003338</concept_id>
       <concept_desc>Information systems~Retrieval models and ranking</concept_desc>
       <concept_significance>500</concept_significance>
       </concept>
   <concept>
       <concept_id>10010147.10010178.10010179</concept_id>
       <concept_desc>Computing methodologies~Natural language processing</concept_desc>
       <concept_significance>300</concept_significance>
       </concept>
 </ccs2012>
\end{CCSXML}
\ccsdesc[500]{Information systems~Retrieval models and ranking}
\ccsdesc[300]{Computing methodologies~Natural language processing}

\keywords{Retrieval-augmented generation, multi-hop retrieval, graph-based retrieval, hypergraphs, attention heads, mechanistic interpretability}

\maketitle

\section{Introduction}
\label{sec:intro}

Retrieval-Augmented Generation (RAG) grounds Large Language Models (LLMs) in external knowledge by retrieving passages from a corpus
and conditioning the answer on them~\cite{rag,ragsurvey}. Retrieval-Augmented Generation (RAG) enables LLMs to leverage the vast and continuously evolving knowledge available on the Web, supporting more accurate and up-to-date question answering. Standard RAG retrieves passages based on their relevance to the original
question, working well for single-hop questions but struggling with multi-hop questions~\cite{hotpotqa,twowiki,musique}. In
multi-hop questions, some essential evidence may not be directly relevant to the original question but is instead connected through
intermediate entities. We call such an intermediate entity a bridge entity, and the question often refers to it only implicitly.
For example, to answer \emph{Where did the author of Princeps Pastorum die?}, a system must find that the author is Pope John~XXIII
and then retrieve his biography, although the question never mentions him. A retriever that compares passages with the question as
a whole finds the passage about Princeps Pastorum but rarely the biography, so the second piece of evidence is often
missed~\cite{ircot,hipporag}.

\begin{figure}[t]
  \centering
  \includegraphics[width=\columnwidth]{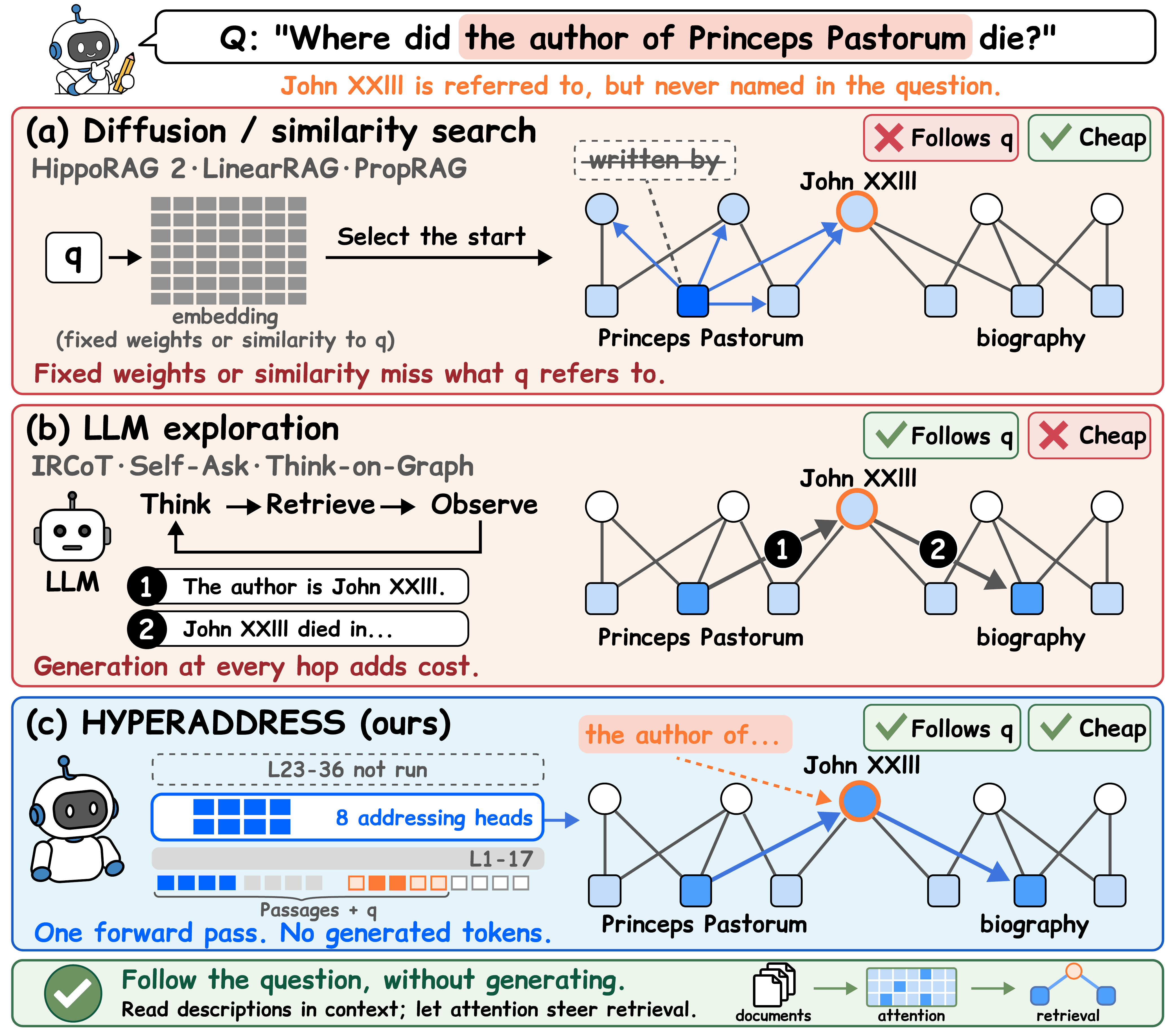}
  \caption{Three paradigms on the same graph: only \mbox{\method} follows the question without generating.}
  \label{fig:paradigms}
\end{figure}

Graph-based RAG (GraphRAG) organizes passages into a graph and retrieves evidence along its edges, enabling the discovery of evidence
passages connected to initially retrieved passages through intermediate entities. While existing systems construct graphs using
different representations~\cite{raptor,graphrag,hipporag2,lightrag,proprag,linearrag}, their multi-hop retrieval strategies
generally fall into two categories (Figure~\ref{fig:paradigms}). Diffusion- and similarity-based methods expand retrieval with a
fixed procedure: HippoRAG~2~\cite{hipporag2} and LinearRAG~\cite{linearrag} spread relevance over the graph with Personalized
PageRank (PPR)~\cite{pagerank,topicpagerank}, and PropRAG~\cite{proprag} searches paths of propositions by their similarity to the
question. Exploration-based methods such as IRCoT~\cite{ircot} and Self-Ask~\cite{selfask} rely on LLM generation for two tasks: identifying intermediate entities to guide retrieval (semantic understanding) and inferring and following relationships among entities (relational exploration). These methods iteratively retrieve passages and generate reasoning based on previously retrieved evidence, while exploring entity relationships through textual contexts.

Both strategies have limitations. (i)~\textbf{Diffusion- and similarity-based methods accumulate relevance errors across hops.} At
each hop, these methods expand retrieval without identifying the entity that the question refers to. For example, PPR propagates
relevance scores using fixed edge weights, regardless of which connections are relevant to the question. As shown in
Figure~\ref{fig:paradigms}(a), the question asks where the author of Princeps Pastorum died but does not explicitly name John~XXIII.
While John~XXIII is the relevant bridge entity, PPR also propagates relevance scores to other connected entities that are
irrelevant to the question. These irrelevant entities can further propagate scores to their neighbors, causing retrieval errors to
accumulate across hops. Similarity-based methods rank passages by their relevance to the original question, often missing essential
evidence at later hops that is connected through intermediate entities. (ii)~\textbf{Exploration-based methods incur high generation costs and underutilize structural information.} First, semantic understanding requires step-by-step LLM generation, where each hop relies on previously retrieved passages and generated reasoning. As the retrieved evidence and reasoning accumulate, the input context grows longer, increasing the computational cost with each hop. Second, relational exploration typically operates over flat textual contexts without explicitly modeling structural relationships among entities, as in graphs. For questions involving complex entity relationships, LLMs must recover and combine relationships scattered across multiple passages. Compared with explicit graph structures, this increases the risk of missing important relationships and may require additional LLM reasoning, further increasing computational costs.

These limitations raise a key question: \textit{Can we combine structural information with question semantics for multi-hop retrieval at low computational cost?}

Our strategy is to decouple semantic guidance from structural exploration, which exploration-based methods typically handle together through LLM generation. Instead of relying on LLMs to generate reasoning at every hop, we use LLMs to provide question-aware semantic signals and graph-based computation to explore structural relationships.

\textbf{Semantic guidance.} The first challenge is whether we can leverage LLMs' semantic understanding capabilities without relying on LLM generation. We revisit how LLMs resolve implicit entity references. Interestingly, an LLM resolves implicit entity references while reading the passages and before generating any token~\cite{latentmultihop,hoppinglate,twohopcontext}. This resolution is concentrated in a handful of attention heads, which we call \emph{addressing heads} (Section~\ref{sec:pointers}). When an LLM reads retrieved passages followed by a multi-hop question, these heads move information from the mentions of each described entity into the words that describe it. For \emph{the author of Princeps Pastorum}, this information comes from the mention of John~XXIII in the passage about Princeps Pastorum. One forward pass does this for the bridges of several hops, each at the words of its own description. The heads can be identified without labels, and ablating them halves the LLM's accuracy on two-hop questions whose bridge is described. In one forward pass, the addressing heads thus score how strongly the question refers to each entity in the passages and to each of its mentions.\\
\textbf{Structural exploration.} The next challenge is how to use these semantic signals to explore relationships among entities without relying on LLM-generated reasoning. To this end, we represent passages and their shared entities in a description-labelled hypergraph and use the addressing scores to steer Personalized PageRank (PPR). The hypergraph explicitly captures structural connections among passages through shared entities, while the addressing scores guide the walk toward entities and mentions relevant to the question. In this way, multi-hop relational exploration is performed through graph-based computation rather than step-by-step LLM generation.
\textbf{\method}, short for \underline{\textbf{Hyper}}graph retrieval via in-context \underline{\textbf{Address}}ing, implements this strategy through two components: an offline description-labelled hypergraph and online attention-steered PageRank.

Our contributions are summarized as follows:
\begin{itemize}
  \item We show that an LLM reading retrieved passages with a multi-hop question identifies the entity that each description refers
  to through a handful of addressing heads. These heads can be identified without labels, and the LLM depends on them when a bridge
  is described. They cover several hops in one forward pass.
  \item We propose \method, a retriever that runs Personalized PageRank on a description-labelled hypergraph and takes both its
  restart distribution and its transition weights from the addressing heads.
  \item On three multi-hop benchmarks, \method outperforms the strongest of nine baselines, PropRAG, by 3.5 points of Recall@5 and
  7.4 points of All@5 on average. Its gains grow with the number of hops, to 6.3 points at four hops on MuSiQue. It generates no
  token at query time, answers more than twice as fast, and builds its index at half the cost.
\end{itemize}

\section{Related Work}
\label{sec:related}

\textbf{Graph-based RAG.} Graph-based RAG systems differ in what their index stores. Hierarchical indexes store summaries:
RAPTOR~\cite{raptor} recursively clusters and summarizes passages into a tree, and Microsoft's GraphRAG~\cite{graphrag} summarizes
the communities of an LLM-extracted entity graph. Knowledge-graph indexes store relations between entities. HippoRAG~\cite{hipporag}
and HippoRAG~2~\cite{hipporag2} extract them as open information extraction triples, and LightRAG~\cite{lightrag} writes
LLM-generated descriptions on entities and edges. G-Retriever~\cite{gretriever} and GFM-RAG~\cite{gfmrag} run graph neural networks
over such graphs. Other indexes store propositions~\cite{proprag} or groups of related facts~\cite{sirerag}, and HopRAG~\cite{hoprag}
connects passages with LLM-generated questions. Relation-free indexes record only which entities co-occur: LinearRAG~\cite{linearrag}
links passages and sentences through the entities they mention, and HGRAG~\cite{hgrag} treats entities as vertices and passages as
hyperedges.

\textbf{Multi-hop traversal.} Multi-hop retrieval methods also differ in what decides each hop. In diffusion-based methods, relevance
spreads along the edges of the graph from the entities and passages that match the question. HippoRAG, HippoRAG~2, and LinearRAG use
Personalized PageRank~\cite{pagerank,topicpagerank}, and HGRAG uses hypergraph diffusion. HippoRAG~2 filters the seed facts with an
LLM, and GFM-RAG learns query-dependent propagation from labelled questions. PropRAG~\cite{proprag} searches proposition paths with a
beam search that scores each step by embedding similarity, without an LLM. In exploration-based methods, an LLM decides each hop.
IRCoT~\cite{ircot} and Self-Ask~\cite{selfask} interleave retrieval with generated reasoning or follow-up questions, and
Think-on-Graph~\cite{tog} and HopRAG let the LLM choose the next node. ChainRAG~\cite{chainrag} rewrites each sub-question with the
entity found at the previous hop, and BridgeRAG~\cite{bridgerag} ranks later-hop candidates with an LLM judge conditioned on the
bridge evidence. EfficientRAG~\cite{efficientrag} trains small models to replace the LLM in this loop.

\textbf{Attention as a retrieval signal.} Attention heads have been used to retrieve information from the context. Retrieval heads
copy information from long inputs~\cite{retrievalhead}. ICR~\cite{icr}, QRHead~\cite{qrhead}, and BlockRank~\cite{blockrank} score
the passages of a prompt by the attention that the query pays to them. LOCOS~\cite{locos} selects heads by their logit contribution
rather than their attention and finds heads that retrieve non-literal matches. The attention of a question-answering reader has also
been distilled into a retriever~\cite{readerretriever}. Interpretability studies describe how models resolve references in context:
through binding IDs~\cite{bindingid}, through lookback mechanisms that retrieve information bound at an earlier
position~\cite{lookback}, or through a mixture of positional and lexical mechanisms~\cite{mixing}. Other studies trace how a model
resolves the bridge entity of a two-hop question~\cite{latentmultihop,hoppinglate,twohopcontext}.

\section{In-Context Addressing in Multi-hop RAG}
\label{sec:pointers}

\begin{figure}[t]
  \centering
  \includegraphics[width=\columnwidth]{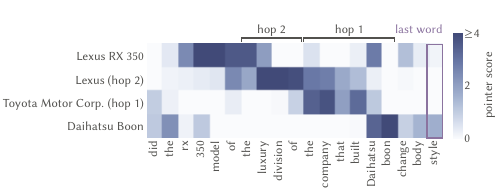}
  \caption{Pointer scores of the addressing heads at each word of a three-hop MuSiQue question. Each bridge scores highest at its own
  description and low at the last word.}
  \label{fig:pointermap}
\end{figure}

An LLM that reads retrieved passages together with a multi-hop question identifies the entity that each description in the
question refers to before it generates anything. Exploration-based retrieval, however, waits for the LLM to generate every hop.
Since a Transformer moves information between positions only through attention, this identification relies on attention
heads~\cite{bindingid,mixing,lookback}. Before introducing our method, we therefore analyze the attention of an LLM while it reads the
passages and the question. The analysis uses Figure~\ref{fig:pointermap}, a three-hop question from MuSiQue~\cite{musique}, as a
running example. Each observation is measured over hundreds of held-out probes or test questions, and Appendix~\ref{app:pre} gives
further results.

\textbf{Analysis Setup.} Our reader is Qwen3-4B~\cite{qwen3}. It reads the top passages of a first-stage retriever followed by the
question in one prompt and generates no token. The embedding model is NV-Embed-v2~\cite{nvembed}. For a head $h$, a
question token $t$, and a candidate entity $e$ whose mentions occupy the positions $M_e$, we measure the share of the head's output
at $t$ that comes from these mentions:
\begin{equation}
  s_h(t,e)=\frac{\big\lVert\sum_{j\in M_e}\alpha_h(t,j)\,W_O^h\,v_h(j)\big\rVert}{\big\lVert\sum_{j}\alpha_h(t,j)\,W_O^h\,v_h(j)\big\rVert},
  \label{eq:share}
\end{equation}
where $\alpha_h(t,j)$ is the attention weight, $v_h(j)$ the value vector, and $W_O^h$ the head's slice of the output projection.
Unlike an attention weight, this share measures how much the mentions contribute to what the head writes at $t$. The \emph{pointer
score} of $e$ is $\max_t s_h(t,e)$, z-normalized over the candidate entities and averaged over a set of heads $H$. We select the
heads without any benchmark label, from probes generated from the index of the corpus. A probe consists of a passage, its nine
nearest passages, and a question built around one description that the passage gives of an entity (Section~\ref{sec:index}); its
target is that entity. The eight of the 1,152 heads with the highest mean reciprocal rank of the target on 600 probes are the
\emph{addressing heads}. Figure~\ref{fig:pointermap} shows their pointer scores at every word of a question.

\begin{table}[t]
  \centering
  \caption{Reader accuracy (\%) after mean ablation of a head set. Probes: 400 held out, gold name not in context. MuSiQue: 304 two-hop
  test questions with two gold passages and four distractors. Random heads: change in points over three draws.}
  \label{tab:causal}
  \footnotesize
  \setlength{\tabcolsep}{4pt}
  \begin{tabular}{lccc}
    \toprule
    & & \multicolumn{2}{c}{MuSiQue, bridge} \\
    \cmidrule(lr){3-4}
    ablated heads & probes & described & named \\
    \midrule
    none & 76.5 & 40.1 & 62.5 \\
    8 addressing heads & 48.5 & 19.7 & 53.6 \\
    \quad at the description only & --- & 27.0 & --- \\
    \quad at as many other tokens & --- & 37.2 & --- \\
    8 random, layers 18--22 & $-2.8$ to $-1.5$ & $-4.0$ to $-2.0$ & $-3.0$ to $-1.3$ \\
    8 random, any layer & $-0.8$ to $+0.8$ & $-1.0$ to $+0.7$ & $-1.3$ to $+1.0$ \\
    \bottomrule
  \end{tabular}
\end{table}

\textbf{Observation 1: Sparse Addressing Heads.} Our first key observation is that a handful of mid-layer heads identify the entity
that a description refers to, and that the reader depends on them. The eight addressing heads all lie in layers 18 to 22 of 36, and
on 400 held-out probes they rank the target first for 59.0\%. Table~\ref{tab:causal} replaces their output by its mean. The reader's
accuracy then falls from 76.5\% to 48.5\% on the probes and halves on two-hop MuSiQue questions with a described bridge, from 40.1\%
to 19.7\%. With the bridge named, it falls by 8.9 points. The reader thus depends on these heads most when it must identify a
described entity (difference of drops 11.5, 95\% CI $[3.9, 19.1]$). Ablating the heads at the words of the description alone costs
13.2 of the 20.4 points, and at as many other tokens 3.0. Eight random heads cost at most 4.0 points from the same layers and 1.3 from
any layer. This sparsity motivates a reader that uses only the addressing heads and stops its forward pass after layer 22, with heads
that each corpus selects from its own probes (Section~\ref{sec:walk}).

\textbf{Observation 2: Word-Level Multi-hop Addressing.} Our second key observation is that one forward pass identifies the bridges
of several hops, each at the words of its own description. Figure~\ref{fig:pointermap} shows a question that describes each bridge
through the one before it. A single embedding of this question mixes the hops, whereas the pointer scores separate them by word. We
place the gold passages of each three- and four-hop MuSiQue question among ten passages and rank the entities that the question does
not name. All bridges of a question are among the five highest-scoring candidates for 79.8\% and 75.6\% of the questions, and each
bridge alone for 85.0\% to 89.8\%. These five candidates set the restart of the walk (Section~\ref{sec:walk}). Each bridge scores
highest inside the words of its own description, the part that no inner description covers, 2.2 to 2.7 times as often as the length
of that part predicts. At the later hops, the highest-scoring entity at these words is the bridge that they describe about twice as
often as chance (64.0\% against 34.0\% at the second hop of three, 53.5\% against 23.0\% at the third of four).

By contrast, scores read only at the last question token, where an answer would be generated and where BlockRank~\cite{blockrank}
reads relevance, rank the later bridges much lower. For the second bridge of three-hop questions, the top-five rate falls from
85.3\% to 52.3\%. For the third bridge of four-hop questions, it falls from 85.0\% to 36.2\%, whereas at the first hop the rates
fall by 9.7 and 11.8 points. With the last token, all bridges are in the top five for 42.3\% and 16.5\% of the questions. Reading
the pointer scores at every word thus gives, in one forward pass, a score for each candidate entity and for each of its mentions.
This motivates a walk that takes both its restart and its transitions from these scores (Section~\ref{sec:walk}).

\begin{figure*}[t]
  \centering
  \includegraphics[width=\textwidth]{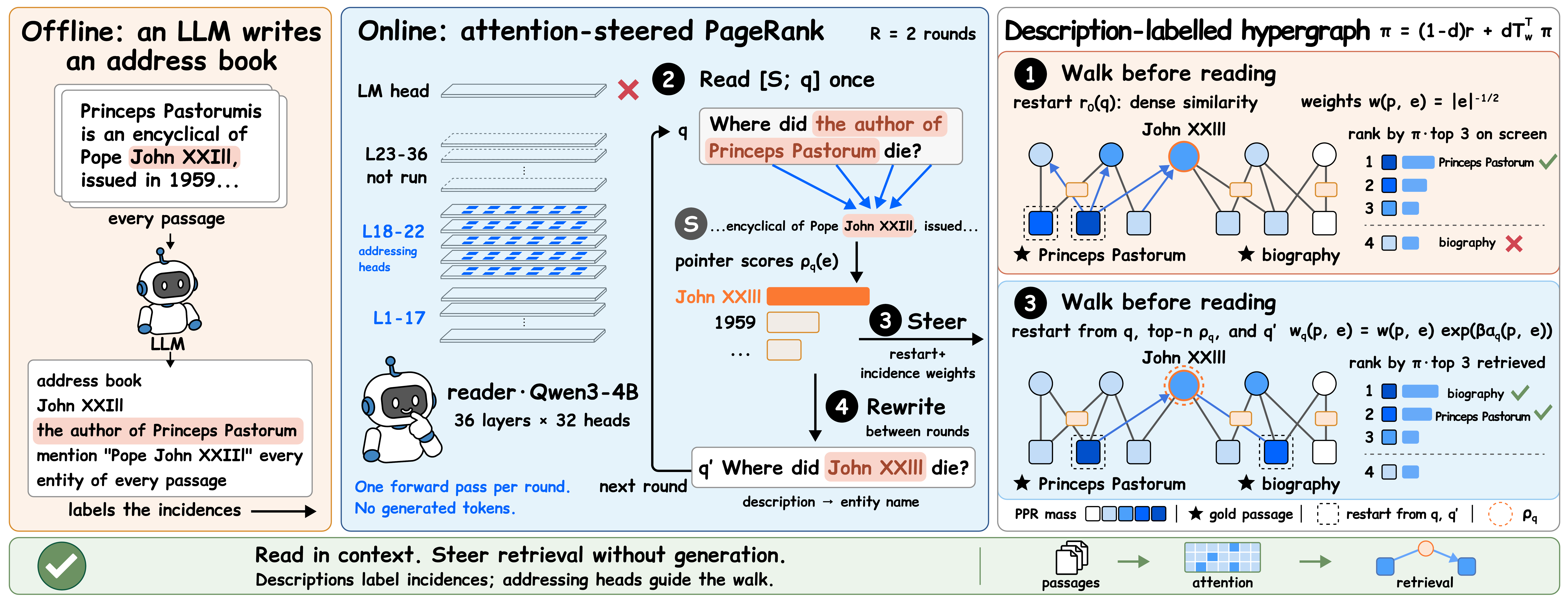}
  \caption{Overview of \method. Offline, an LLM labels each incidence with descriptions. Online, each of $R$ rounds walks
  (\ding{192}), reads the passages in one forward pass (\ding{193}), steers restart and transitions (\ding{194}), and
  rewrites one description into a name (\ding{195}).}
  \label{fig:overview}
\end{figure*}

\textbf{Observation 3: Context-Dependent Addressing.} Our third key observation is that this identification depends on reading the
passages in context. An index encodes each passage once, apart from any question, which is what lets it scale.
Table~\ref{tab:reading} tests whether the identification survives such encoding. In context, the addressing heads rank the target
first for 59.0\% of the probes and 55.3\% of the MuSiQue bridges, and the embeddings for at most 37.0\% and 34.2\%. The paired gains
in mean reciprocal rank are .174 and .198 on the probes (95\% CIs above 0). With either side encoded apart (the question, or each
passage, as in caches of precomputed passages~\cite{blockattn,memory3,tome}), the same heads fall below both embeddings. With both
apart, their mean reciprocal rank falls by .403 and .383 (95\% CIs above 0), and they keep 29\% of their margin over chance on both
sets. Position does not explain the loss. Every encoding keeps the token positions of the joint prompt, and removing the rotary
position encoding from the heads changes their mean reciprocal rank by less than .01 (Appendix~\ref{app:pre}). End to end, scoring
entities by the similarity of the question to their stored descriptions recovers 1.8 of the 8.4 points that the addressing heads add
(Section~\ref{sec:gain}). Because the identification depends on context, \method computes it at query time and stores only
descriptions in the index (Section~\ref{sec:index}).

\begin{table}[t]
  \centering
  \caption{Mean reciprocal rank (top-1 \%) of the target among the candidate entities. MuSiQue: bridges whose gold passage is not in the
  walk's top five.}
  \label{tab:reading}
  \footnotesize
  \setlength{\tabcolsep}{4pt}
  \begin{tabular}{lcc}
    \toprule
    & probes (400) & MuSiQue (304) \\
    \midrule
    \multicolumn{3}{l}{\emph{read in context}} \\
    \oursrow
    \quad addressing heads & .734 (59.0) & .705 (55.3) \\
    \multicolumn{3}{l}{\emph{encoded apart}} \\
    \quad embedding, question and name & .560 (37.0) & .529 (34.2) \\
    \quad embedding, question and passages & .537 (34.0) & .502 (30.9) \\
    \quad addressing heads, question apart & .482 & .440 \\
    \quad addressing heads, passages apart & .512 & .470 \\
    \quad addressing heads, both apart & .332 & .323 \\
    \midrule
    chance & .169 & .160 \\
    \bottomrule
  \end{tabular}
\end{table}

\section{Attention-Steered Hypergraph Retrieval}
\label{sec:method}

Based on these observations, \method works in two stages (Figure~\ref{fig:overview}). Offline, it organizes the
corpus into a \emph{description-labelled hypergraph}, which stores each relation on the incidence between a passage and an entity
(Section~\ref{sec:index}). Online, it retrieves with \emph{Attention-Steered PageRank}, a random walk on this hypergraph whose restart
and transition weights are set by the addressing heads of the reader (Section~\ref{sec:walk}).

\textbf{Problem setting.} Given a corpus $\mathcal{P}$ of passages and a question $q$, a retriever ranks the passages so that the
evidence for $q$ appears among the first $k$. For a multi-hop question, part of this evidence is about entities to which $q$ refers
only through descriptions.

\subsection{Offline: Description-Labelled Hypergraph}
\label{sec:index}

\textbf{Definition.} A description-labelled hypergraph $\mathcal{H}=(\mathcal{P},\mathcal{E},w,D)$ has the passages as its vertices
and a set $\mathcal{E}$ of hyperedges, one per entity, each containing the passages that mention its entity. A pair $(p,e)$ with
$p\in e$ is an \emph{incidence}; it has a weight $w(p,e)>0$ and a label $D(p,e)$: the descriptions by which passage $p$ identifies
entity $e$ without naming it. A relation is thus stored as the label of the incidence between the passage that states it and the
entity that it concerns, rather than as an edge between two entities.

\textbf{Construction.} An LLM reads every passage once and writes its \emph{address book}. For each entity that the passage
mentions, an entry gives the canonical name of the entity, its mentions in the passage, and one to three descriptions. Each
description identifies the entity through a fact stated in the passage and is phrased the way a question would refer to it.
Canonical names and their aliases group the entries of all passages into hyperedges. Each hyperedge also contains every passage that
names its entity, and the descriptions label its incidences. An incidence has the weight $w(p,e)=|e|^{-1/2}$, so that the walk moves
less probability through an entity that many passages mention. The LLM also writes the facts that the descriptions leave out and a
question that the passage answers. Together with the passage's own sentences and one card per entity, which lists the entity's name
with its descriptions, these form the \emph{content items} of the passage.

\textbf{Design rationale.} A knowledge graph makes entities its nodes and extracted relations its edges. $\mathcal{H}$ inverts these
roles: passages are the vertices, entities the hyperedges, and what a passage says about an entity stays on their incidence, in the
words of the passage. This placement serves both stages. Offline, a label is phrased the way a question refers to an entity. The
initial restart can therefore match a description and bring the passage that states the relation into the reader's context. Online,
the reader identifies the entity from the text of that passage at query time (Observation~3). Since two passages are connected only
through an entity, the reader's scores re-weight every transition out of a passage that it has read. A new passage is added with its
own address book and joins the hyperedges of the entities that it names. No other passage is processed again, so the index supports
non-parametric continual learning~\cite{hipporag2}.

\subsection{Online: Attention-Steered PageRank}
\label{sec:walk}

\method ranks the passages with Personalized PageRank on $\mathcal{H}$ and lets the reader set both of its inputs, the restart
distribution and the transition weights, over $R$ rounds (Algorithm~\ref{alg:aspr}).

\begin{algorithm}[t]
  \caption{Attention-Steered PageRank}
  \label{alg:aspr}
  \small
  \begin{algorithmic}[1]
    \Require hypergraph $\mathcal{H}=(\mathcal{P},\mathcal{E},w,D)$, question $q$, reader with addressing heads, rounds $R$, passages per round $m$
    \Ensure a ranking of $\mathcal{P}$
    \State $q_0\gets q$;\; $r\gets r_0(q)$;\; $w_q\gets w$
    \State $\pi\gets\mathrm{PPR}(r,w_q)$ \Comment{Eq.~\ref{eq:ppr}}
    \For{$t=1,\dots,R$}
      \State $S\gets$ the first $m$ passages of $\mathrm{Rank}(\pi)$
      \State $(\rho_q,a_q)\gets\mathrm{Read}(S,q)$ \Comment{one forward pass}
      \State steer the restart $r$ with $\rho_q$ \Comment{Eq.~\ref{eq:restart}}
      \State steer the weights $w_q$ of the incidences on $S$ with $a_q$ \Comment{Eq.~\ref{eq:edges}}
      \If{$t<R$} \Comment{between rounds}
        \State $q\gets\mathrm{Rewrite}(q)$;\; add $r_0(q)$ to $r$
      \EndIf
      \State $\pi\gets\mathrm{PPR}(r,w_q)$
    \EndFor
    \State \Return $\mathrm{Rank}(\pi)$
  \end{algorithmic}
\end{algorithm}

\textbf{Walk.} The walk alternates between passages and hyperedges: from a passage $p$ it moves to a hyperedge $e\ni p$ with
probability proportional to $w(p,e)$, and from a hyperedge $e$ to a passage $p\in e$ with probability proportional to $w(p,e)$. This
is a random walk on a hypergraph with edge-dependent vertex weights~\cite{zhouhyper,hyperwalk}. Given a restart distribution $r$
over passages and hyperedges and the transition matrix $T_w$ of the walk, the stationary scores satisfy
\begin{equation}
  \pi=(1-d)\,r+d\,T_w^{\top}\pi,
  \label{eq:ppr}
\end{equation}
with damping factor $d$. The initial restart $r_0(q)$ is set before any reading. It places probability on every passage in
proportion to its dense similarity to $q$, on the incidences whose labels are most similar to $q$, and on the passages of the
content items most similar to $q$. $\mathrm{Rank}(\pi)$ orders the passages by their stationary score, fused with their dense
similarity to the original question $q_0$.

\textbf{In-context addressing.} In each round, the reader processes the set $S$ of the $m$ top-ranked passages, in their original
text, followed by the question. It does so in one forward pass that stops after the last layer with an addressing head and generates
no token. The candidates are the entities that have an incidence with a passage in $S$ and that $q$ does not name. For each
candidate $e$, the heads give its pointer score $\rho_q(e)$, computed from Equation~\eqref{eq:share} as in
Section~\ref{sec:pointers}. For each incidence $(p,e)$ with $p\in S$, the same score restricted to the mentions of $e$ inside $p$
gives its incidence score $a_q(p,e)$.

\textbf{Steering the restart.} The $n$ candidates with the highest pointer scores receive a share $\lambda$ of the restart
probability, split between their hyperedges and the passages in $S$ that mention them:
\begin{equation}
  r\leftarrow(1-\lambda)\,r+\lambda\!\!\sum_{e\in\mathrm{top}_n(\rho_q)}\!\!\mathrm{softmax}(\rho_q)_e\,\tfrac12\big(\mathbf{1}_e+u_e\big),
  \label{eq:restart}
\end{equation}
where $\mathbf{1}_e$ is the indicator of hyperedge $e$ and $u_e$ the uniform distribution over the passages in $S$ that mention $e$.
The walk then restarts both from the passages similar to the question and from the entities that the question refers to.

\textbf{Steering the transitions.} Every incidence that the reader has read is re-weighted by its incidence score, in both directions:
\begin{equation}
  w_q(p,e)=w(p,e)\,\exp\big(\beta\,a_q(p,e)\big),
  \label{eq:edges}
\end{equation}
and $T_{w_q}$ replaces $T_w$ in Equation~\ref{eq:ppr}. A passage keeps the incidence scores of the last round in which it was read.
A passage in $S$ thus passes more of its probability to the entities that the question refers to, instead of distributing it by the
fixed weights.

\textbf{Rounds and cost.} Between rounds, \method replaces one description in the question with the name of the entity that it most
likely refers to, selected by the pointer scores and the labels (Appendix~\ref{app:impl}). The dense terms of the restart can then
match this name. The rewritten question adds its own restart and is the question read in the next round. The reader sees the
passages as plain text. Appending the labels of their incidences does not change Recall@5 significantly (82.8 against 83.1), because
each label is drawn from a passage that the reader already reads. \method runs two rounds. All generation happens offline, when the
index is built. At query time, \method encodes the question and, in each round, runs one forward pass of the reader and one walk on
the hypergraph. Section~\ref{sec:cost-exp} measures the cost of both stages.

\begin{table*}[t]
  \centering
  \caption{Retrieval results (\%) on the test questions. All@5: questions with all gold passages in the top five. Gen.: generates
  text at query time. Best in bold, second best underlined.}
  \label{tab:main}
  \small
  \setlength{\tabcolsep}{4.6pt}
  \begin{tabular}{lcccccccccccc}
    \toprule
    & & \multicolumn{3}{c}{MuSiQue} & \multicolumn{3}{c}{2Wiki} & \multicolumn{3}{c}{HotpotQA} & \multicolumn{2}{c}{Average} \\
    \cmidrule(lr){3-5}\cmidrule(lr){6-8}\cmidrule(lr){9-11}\cmidrule(lr){12-13}
    Method & Gen. & R@2 & R@5 & All@5 & R@2 & R@5 & All@5 & R@2 & R@5 & All@5 & R@5 & All@5 \\
    \midrule
    \multicolumn{13}{l}{\emph{Dense retrieval}} \\
    NV-Embed-v2 & \ding{55} & 52.9 & 69.9 & 38.9 & 69.1 & 76.8 & 51.1 & 84.0 & 94.8 & 89.8 & 80.5 & 59.9 \\
    \multicolumn{13}{l}{\emph{Graph-based RAG}} \\
    RAPTOR & \ding{55} & 41.7 & 59.4 & 35.4 & 60.8 & 67.8 & 61.4 & 72.3 & 83.6 & 78.5 & 70.3 & 58.4 \\
    HippoRAG~2 & \ding{51} & 56.1 & 75.3 & 48.5 & 77.3 & 93.2 & 81.5 & 80.4 & 95.8 & 92.5 & 88.1 & 74.2 \\
    PropRAG & \ding{55} & 57.5 & 77.8 & \underline{53.3} & 71.1 & \underline{94.4} & 83.6 & 84.9 & \underline{96.3} & \underline{92.9} & \underline{89.5} & \underline{76.6} \\
    HGRAG & \ding{51} & 53.1 & 72.9 & 48.5 & 71.2 & 91.4 & 79.4 & 82.3 & 95.7 & 91.6 & 86.7 & 73.2 \\
    LinearRAG & \ding{55} & 54.7 & 73.8 & 48.2 & 75.8 & 92.2 & 85.3 & 81.1 & 96.0 & 92.2 & 87.3 & 75.2 \\
    \multicolumn{13}{l}{\emph{Exploration-based retrieval}} \\
    IRCoT (Qwen3-4B) & \ding{51} & 56.1 & 76.4 & 47.8 & 77.0 & 90.8 & 82.8 & 84.3 & 94.0 & 90.1 & 87.1 & 73.6 \\
    IRCoT (Qwen3.5-27B) & \ding{51} & \underline{58.3} & \underline{78.9} & 49.6 & \underline{78.1} & 92.5 & \underline{86.1} & \underline{85.0} & 94.8 & 91.8 & 88.7 & 75.8 \\
    \multicolumn{13}{l}{\emph{Attention-based re-ranking}} \\
    ICR (Qwen3-4B) & \ding{55} & 48.1 & 67.1 & 36.5 & 66.8 & 75.3 & 48.6 & 79.8 & 92.2 & 85.1 & 78.2 & 56.7 \\
    QRHead (Qwen3-4B) & \ding{55} & 53.3 & 74.4 & 48.0 & 70.0 & 81.8 & 59.1 & 82.4 & 95.9 & 92.5 & 84.0 & 66.5 \\
    \midruleours
    \oursrowlast
    \method (Qwen3-4B) & \ding{55} & \textbf{61.7} & \textbf{83.1} & \textbf{62.0} & \textbf{80.6} & \textbf{97.5} & \textbf{92.8} & \textbf{87.3} & \textbf{98.4} & \textbf{97.1} & \textbf{93.0} & \textbf{84.0} \\
    \bottomruleours
  \end{tabular}
\end{table*}

\section{Experiments}
\label{sec:experiments}

We evaluate \method along five research questions. \textbf{RQ1}: does it retrieve the evidence better than
diffusion- and exploration-based methods? \textbf{RQ2}: where does its gain come from? \textbf{RQ3}: what does it cost to build and
to query? \textbf{RQ4}: do better passages yield better answers? \textbf{RQ5}: does it work with other readers?

\subsection{Setup}
\label{sec:setup}

\textbf{Datasets and metrics.} We follow the protocol of HippoRAG~2~\cite{hipporag2} on three multi-hop benchmarks:
MuSiQue~\cite{musique}, whose questions take two to four hops, 2WikiMultiHopQA (2Wiki)~\cite{twowiki}, and HotpotQA~\cite{hotpotqa}.
Each benchmark provides 1,000 questions and a corpus of 11,656, 6,119, and 9,811 passages, respectively, which every method indexes as
a whole. The first 200 questions of each benchmark form its development set, and the remaining 800 its test set. Every design choice
and constant of \method was fixed on the development questions of MuSiQue and kept for all benchmarks. Baselines run with their
released configurations. All results are on the test questions; Appendix~\ref{app:all1000} reports all 1,000 questions, where our
reproductions match the published numbers. Recall@$k$ ($k=2,5$) is the fraction of gold passages among the top $k$, and All@5 the
fraction of questions whose gold passages all appear in the top five. For question answering, Qwen3.5-27B answers every question from
the top five passages of each retriever, and we report exact match (EM) and F1. Differences are tested over questions with a paired
bootstrap at the 95\% level.

\begingroup\setlength{\emergencystretch}{1.5em}
\textbf{Baselines.} We compare \method with four families of retrievers. (i)~Dense retrieval: NV-Embed-v2~\cite{nvembed}.
(ii)~Graph-based RAG: RAPTOR~\cite{raptor} retrieves over a tree of recursive summaries; HippoRAG~2~\cite{hipporag2} walks an
LLM-extracted knowledge graph with Personalized PageRank from LLM-filtered seeds; PropRAG~\cite{proprag} searches paths of
LLM-extracted propositions with an LLM-free beam search; HGRAG~\cite{hgrag} diffuses relevance over a hypergraph of entities and
passages; and LinearRAG~\cite{linearrag} propagates relevance over a relation-free graph of entities, sentences, and passages.
(iii)~Exploration: IRCoT~\cite{ircot} interleaves retrieval with chain-of-thought generation, with at most four steps on MuSiQue,
two on the other benchmarks, and ten passages per step, as in HippoRAG~\cite{hipporag}. (iv)~Attention-based re-ranking:
ICR~\cite{icr} and QRHead~\cite{qrhead} re-rank the top twenty passages of the dense retriever by the attention that the question
pays to them, both with Qwen3-4B, the reader of \method. Every baseline uses NV-Embed-v2 and, wherever it needs an LLM,
Qwen3.5-27B~\cite{qwen35}, the LLM that writes the index of \method. All run with their official code, except HGRAG, which releases
none and which we reimplement from its paper (Appendix~\ref{app:baselines}). We also run IRCoT with Qwen3-4B to compare generation
and reading on the same model.\par
\endgroup

\textbf{Implementation.} Qwen3.5-27B writes the index. The reader is Qwen3-4B~\cite{qwen3}, whose eight addressing heads are selected
per corpus from 600 synthetic probes (Section~\ref{sec:pointers}). NV-Embed-v2 encodes passages, descriptions, content items, and
questions. All experiments run on one NVIDIA RTX PRO 6000 GPU (96\,GB).

\subsection{RQ1: Retrieval Effectiveness}
\label{sec:main}

Table~\ref{tab:main} reports the retrieval results. \method ranks first on every benchmark and every metric. Averaged over the three
benchmarks, it improves Recall@5 by 3.5 points and All@5 by 7.4 points over the strongest baseline, PropRAG. Every improvement over
the strongest baseline of its column is significant (paired bootstrap, 95\%). \method also ranks first on the single-hop questions of
NQ and PopQA (Appendix~\ref{app:singlehop}). We highlight three results.

\textbf{Gains grow with question depth.} The margin over PropRAG is widest on MuSiQue, whose questions take up to four hops (5.3
points of Recall@5 and 8.7 of All@5), and narrowest on HotpotQA, whose questions take two (2.1 and 4.2). Within MuSiQue, the margin
also grows with the number of hops (Section~\ref{sec:gain}).

\textbf{More questions get complete evidence.} On every benchmark, the gain in All@5 exceeds the gain in Recall@5: \method retrieves
not only more of the passages that a question needs, but also more often all of them. On 2Wiki, 92.8\% of the questions have their
complete evidence in the top five, against 86.1\% with IRCoT on the 27B model, the strongest baseline on this metric.

\textbf{Steering outperforms reasoning and re-ranking.} IRCoT and \method can run on the same 4B model. IRCoT generates a reasoning
step for every hop, whereas \method reads the passages without generating. With this model, \method is ahead by 5.9 points of
Recall@5 and 10.4 of All@5 on average, and a 27B model does not close the gap. The same reader can re-rank the dense passages by
attention, as ICR and QRHead do. \method is ahead of the better of them by about 9 points of Recall@5 and 17 of All@5, because
re-ranking reorders the passages, whereas steering reaches the passages that the dense retriever misses.

\subsection{RQ2: Where the Gain Comes From}
\label{sec:gain}
\begin{figure}[t]
  \centering
  \includegraphics[width=\columnwidth]{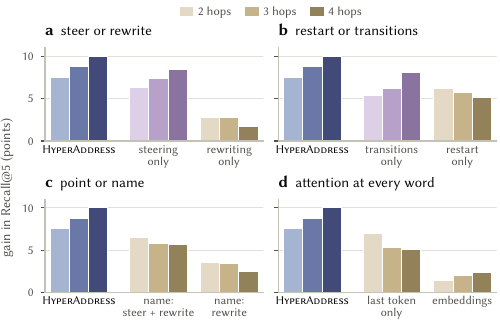}
  \caption{Gain in Recall@5 over the walk without reading, by number of hops (MuSiQue test). Indigo and lilac: the addressing heads set the
  transitions from every word.}
  \label{fig:hops}
\end{figure}

Figure~\ref{fig:hops} decomposes the gain of \method on MuSiQue, the benchmark with the deepest questions, by number of hops, and
Table~\ref{tab:gain} in Appendix~\ref{app:tables} lists every variant. The gain grows with depth when the addressing heads, read at
every word of the question, set the transitions of the walk.

\textbf{Steering adds 8.4 points.} Without reading, the walk on the hypergraph reaches 74.7 points of Recall@5, close to HippoRAG~2
(75.3) and LinearRAG (73.8). A walk with fixed transitions thus reaches a similar level whether its graph stores triples,
co-oc\-cur\-rences, or descriptions. Steering the walk with the addressing heads adds 8.4 points, and the gain grows with the number
of hops: 7.6 points at two hops, 8.8 at three, and 10.1 at four. The margin over IRCoT with the 27B model, the strongest baseline on
MuSiQue, grows from 3.6 points at two hops to 6.7 at four. Over PropRAG it grows from 4.5 to 6.3, and over HippoRAG~2 from 6.0 to
11.6.

\textbf{Most of the gain comes from steering.} One forward pass of the reader, with no generation and no rewrite, raises Recall@5
from 74.7 to 80.1 (Table~\ref{tab:readers}), 5.4 of the 8.4 points. A second round of steering reaches 81.7, and rewriting the
question between the two rounds reaches 83.1. Rewriting alone, without steering, reaches 77.3, and its gain does not grow with depth
(Figure~\ref{fig:hops}a). At four hops it adds 1.8 points to the walk, whereas steering alone adds 8.5. A rewrite names one entity per
round, as in ChainRAG~\cite{chainrag}. One forward pass, in contrast, scores the bridges of several hops (Observation~2), and steering
passes all of these scores to the walk. A third round adds 2.9 points at four hops, where a third bridge remains, and less than one
point at two and three hops.

\textbf{Both inputs of the walk contribute.} The steered restart receives 0.02 of the restart probability, on at most five entities
and the passages in $S$ that mention them. Without it ($\lambda=0$), Recall@5 drops by 2.3 points, similarly at every depth (2.2, 2.6,
and 2.0). Without the steered transitions ($\beta=0$), it drops by 2.4 points: 1.3 at two hops, 3.0 at three, and 4.9 at four. With
steered transitions alone, the gain still grows with the number of hops; with a steered restart alone, it shrinks
(Figure~\ref{fig:hops}b). The restart thus helps at every depth, and the transitions help more with every additional hop.

\textbf{Pointer scores outperform a generated name.} Qwen3-4B can instead generate the name of the bridge from the same passages, and
the name can replace the pointer scores (Appendix~\ref{app:impl}). Rewriting the question with the generated name reaches 78.1 points
of Recall@5, 5.0 below \method and 7.5 below at four hops. Using the name for both rewriting and steering reaches 80.9, and the gap
to \method grows with the number of hops (Figure~\ref{fig:hops}c), to 53.5 against 57.9 at four. A name selects one entity, whereas
one forward pass scores every candidate entity and every incidence in $S$, and steering uses all of these scores. Steering with the
pointer scores alone, without any rewrite, already reaches 81.7.

\textbf{Steering outperforms re-ranking the walk.} The reader can also re-rank the passages that the walk finds. With the same
Qwen3-4B, QRHead re-ranks the walk's top twenty passages to 80.0, 96.0, and 97.1 points of Recall@5 on MuSiQue, 2Wiki, and HotpotQA,
below the 83.1, 97.5, and 98.4 of \method. ICR falls below the walk itself on all three. Re-ranking reorders the passages that the
walk has already found, whereas steering changes which passages the next walk reaches.

\textbf{The scores need every word and the addressing heads.} Scoring only the last question token, where an answer would start,
loses 2.2 points, and 4.9 at four hops (Figure~\ref{fig:hops}d). Later bridges score highest at the words of their own descriptions,
not at the last token (Observation~2). Scoring the entities by the similarity between the question and their descriptions, instead
of by the addressing heads, loses 6.6 of the 8.4 points, because the identification depends on context (Observation~3). The choice
of heads also matters. Heads of the same reader selected as in QRHead, or by the logit contribution of LOCOS~\cite{locos}, steer the
same walk over the same candidates and keep 2.4 and 1.2 of the 8.4 points. Both head sets still improve the walk at two hops, where
QRHead reaches 88.0 against 84.7. At four hops, they fall to the level of the walk without reading (50.0 and 47.8, against 47.8),
whereas \method reaches 57.9. Heads selected for query relevance thus help on the first hop, but later hops need heads that identify
the entity each description refers to.

\textbf{The labels set where the walk starts.} The labels enter retrieval through the initial restart. Removing them from the
restart, with the rest of the index unchanged, lowers Recall@5 by 3.2 points, and the loss grows with depth, from 1.3 points at two
hops to 9.1 at four. The wording of the labels matters. Replacing each label with the name of its entity, so that an incidence
only records that a passage mentions the entity, costs 2.2 points. The rewrite uses the labels: selecting the span to rewrite
without them costs 1.1 points.

\textbf{The address book matters once the walk is steered.} We replace the address book with the entities that spaCy recognizes
in each passage, without descriptions or content items, and keep the reader and its heads. Without reading, the two indexes perform
alike (74.2 with spaCy, 74.7 with the address book). After one round of steering, the address book is ahead by 4.2 points (80.1
against 75.9, Table~\ref{tab:readers}), and raises the gain of one round from 1.7 to 5.4 points.

\subsection{RQ3: Indexing and Query Cost}
\label{sec:cost-exp}

\begin{figure}[t]
  \centering
  \includegraphics[width=\columnwidth]{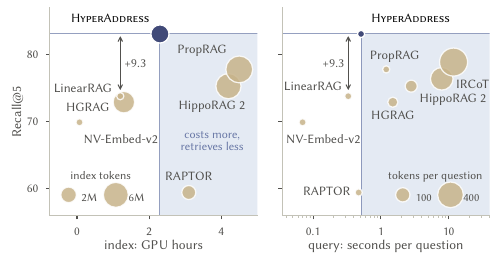}
  \caption{Recall@5 on MuSiQue against indexing time (left) and query time (right, log scale) on one GPU. Marker area: generated tokens.
  IRCoT retrieves from the dense index.}
  \label{fig:cost}
\end{figure}

Figure~\ref{fig:cost} plots the Recall@5 of each method on MuSiQue against its cost, once to build its index and then for every
question; Table~\ref{tab:cost} in Appendix~\ref{app:tables} lists every cost. \method builds its index, head selection included, with
less than half the tokens that HippoRAG~2 and PropRAG generate and in about half their GPU time. At query time, it generates no token
and answers a question in 0.51 seconds. PropRAG takes more than twice as long, HippoRAG~2 five times as long, and IRCoT with the 27B
model 23 times as long. With the same 4B model, IRCoT takes 15 times as long and retrieves 6.7 points less Recall@5. Every cheaper
index or faster method scores at least 9 points of Recall@5 below \method.

\subsection{RQ4: Question Answering}
\label{sec:qa}

\begin{table}[t]
  \centering
  \caption{Question answering (EM/F1, \%) by Qwen3.5-27B from the top five passages of each retriever.}
  \label{tab:qa}
  \footnotesize
  \setlength{\tabcolsep}{3.5pt}
  \begin{tabular}{lcccc}
    \toprule
    Retriever & MuSiQue & 2Wiki & HotpotQA & Average \\
    \midrule
    NV-Embed-v2 & 35.9/46.6 & 54.6/60.4 & 63.5/76.6 & 51.3/61.2 \\
    HippoRAG~2 & 41.1/53.2 & 64.2/72.4 & 64.4/78.1 & 56.6/67.9 \\
    PropRAG & 42.7/53.5 & 65.0/71.4 & 64.9/78.3 & 57.5/67.7 \\
    LinearRAG & 41.0/51.5 & 64.8/70.8 & 65.1/77.8 & 57.0/66.7 \\
    IRCoT (4B) & 41.3/50.9 & 63.2/70.8 & 64.0/77.5 & 56.2/66.4 \\
    IRCoT (27B) & 42.9/55.1 & 64.7/73.1 & 64.8/78.3 & 57.5/68.8 \\
    \midruleours
    \oursrowlast
    \method & \textbf{46.5}/\textbf{59.1} & \textbf{66.8}/\textbf{75.1} & \textbf{67.1}/\textbf{80.1} & \textbf{60.1}/\textbf{71.4} \\
    \bottomruleours
  \end{tabular}
\end{table}

Better retrieval gives better answers. In Table~\ref{tab:qa}, Qwen3.5-27B answers every question from the top five passages
of each retriever. With the passages of \method, it gives the best answers on all benchmarks. On average, it is 2.6 points of
EM above PropRAG and 2.6 of F1 above IRCoT with the 27B model, the strongest baselines on these metrics. On MuSiQue, whose questions
are the deepest, it is 3.6 and 4.0 points above IRCoT with the 27B model.

\subsection{RQ5: Other Readers}
\label{sec:readers}

\begin{table}[t]
  \centering
  \caption{Recall@5 (\%) on MuSiQue with other readers, each with eight addressing heads selected without labels. Walk without reading:
  74.7. 1 round: steering once, no rewriting.}
  \label{tab:readers}
  \footnotesize
  \setlength{\tabcolsep}{4pt}
  \begin{tabular}{lccc}
    \toprule
    Reader & heads (layers) & 1 round & \method \\
    \midrule
    Qwen3-1.7B & 16--17 of 28 & 79.3 & 81.5 \\
    Qwen3-4B & 18--22 of 36 & 80.1 & 83.1 \\
    Qwen3-8B & 18--23 of 36 & 80.5 & 83.5 \\
    Llama-3.1-8B-Instruct & 13--14 of 32 & 80.5 & 83.5 \\
    \bottomrule
  \end{tabular}
\end{table}

We find addressing heads in every reader that we test. We select eight heads for Qwen3-1.7B, Qwen3-8B, and
Llama-3.1-8B-Instruct~\cite{llama3} with the same label-free probes and keep the rest of \method unchanged (Table~\ref{tab:readers}).
Steering improves the walk with every reader. With 1.7B parameters, \method reaches 81.5 points of Recall@5, 2.6 above IRCoT with a
27B model, the strongest baseline. The gain over the walk grows with the reader, from 6.8 points with 1.7B parameters to 8.8 with
either 8B model. In all four models, the selected heads lie in the middle layers.

\section{Conclusion}
\label{sec:conclusion}

We presented \method, a GraphRAG framework in which the attention of an LLM steers a graph walk without generation. The LLM reads the
top passages together with the question in one forward pass, and its addressing heads set both the restart and the transitions of
Personalized PageRank on a description-labelled hypergraph. Without generating any token at query time, \method outperforms nine
baselines on three multi-hop benchmarks, with gains that grow with the number of hops. On these benchmarks, steering a graph walk with
the attention of an LLM retrieves more evidence than letting the LLM generate every hop, in a fraction of the time.

\bibliographystyle{ACM-Reference-Format}
\bibliography{refs}

\appendix
\section{Implementation Details}
\label{app:impl}

\textbf{Constants.} Table~\ref{tab:hparams} lists the constants of \method. They were chosen on the development questions of
MuSiQue and are kept for all benchmarks and readers. NV-Embed-v2 encodes each question twice: with its retrieval instruction against
passages and content items, and with an addressing instruction against labels.

\begin{table}[h]
  \centering
  \caption{Constants of \method.}
  \label{tab:hparams}
  \footnotesize
  \setlength{\tabcolsep}{3pt}
  \begin{tabular}{lll}
    \toprule
    component & constant & value \\
    \midrule
    hypergraph & incidence weight & $|e|^{-1/2}$ \\
    walk & damping $d$; iterations & 0.85; 30 \\
    restart $r_0$ & dense prior & $0.05\times$ min--max similarity \\
    & top labels; top items & 10; 30 \\
    ranking & fusion & $z(\log\pi)+0.5\,z(\text{dense})$ \\
    reading & passages per round $m$; heads & 10; 8 \\
    steer restart & $n$; $\lambda$; temperature & 5; 0.02; 1 \\
    steer transitions & $\beta$ & 1 \\
    rounds & $R$; span threshold; weight & 2; 0.7; 2 \\
    \bottomrule
  \end{tabular}
\end{table}

\textbf{Rewriting.} We select the entity with the highest sum of its z-normalized pointer score and the z-normalized similarity
between the question and its best label in $S$. The candidate spans of the question are its word n-grams of at most 16 words that
start at a determiner or at a capitalized word. The span most similar to the selected entity's best label is replaced by the entity's
name if their cosine similarity is at least 0.7. The restart of the rewritten question is added, with weight 2, to the restarts of
the earlier versions of the question.

\textbf{Generated names.} In the variants of Table~\ref{tab:gain} that generate the bridge name, Qwen3-4B reads the same passages and
question as \method, followed by an instruction to name the entity that the question describes without naming it. It decodes up to 16
tokens, greedily and without thinking mode, as the baselines do (Appendix~\ref{app:baselines}). The name is mapped to the candidate in
$S$ whose name is most similar to it under NV-Embed-v2, so that it selects an entity in every round, as the pointer scores do. This
entity replaces the one selected for rewriting. To steer, it receives the whole share $\lambda$ of the restart, and each of its
incidences in $S$ receives the highest incidence score of that round.

\subsection{Addressing Heads: Additional Results}
\label{app:pre}

\textbf{Binding.} To test whether the addressing heads follow the text or the name, we build 300 probes whose passage also mentions
an entity of the same type as the target, and we edit the passage. When the passage binds the description to a made-up name, the
made-up name becomes the top candidate for 25\% of the probes. When the made-up name replaces the other entity, which the description
does not refer to, it becomes the top candidate for 3\%. The heads thus rank first the entity that the text binds to the description,
whatever its name.

\textbf{Positional rotation.} Removing the rotary position encoding from the queries and keys of the addressing heads changes the
mean reciprocal rank of the target in Table~\ref{tab:reading} by less than .01 (.728 instead of .734 on probes, .702 instead of .705
on MuSiQue). The match is therefore computed from content rather than position.

\begin{table}[t]
  \centering
  \caption{Recall@5 (\%) on MuSiQue test, overall and by number of hops (415, 258, and 127 questions). $\Delta$: change of \method from the
  walk and of each indented row from \method. All are significant (paired bootstrap, 95\%) except the appended labels.}
  \label{tab:gain}
  \footnotesize
  \setlength{\tabcolsep}{3.5pt}
  \begin{tabular}{lccccc}
    \toprule
    & & \multicolumn{3}{c}{by hops} & \\
    \cmidrule(lr){3-5}
    Method & R@5 & 2 & 3 & 4 & $\Delta$ \\
    \midrule
    NV-Embed-v2 & 69.9 & 78.2 & 69.6 & 43.5 &  \\
    HippoRAG~2 & 75.3 & 86.3 & 71.8 & 46.3 &  \\
    PropRAG & 77.8 & 87.8 & 74.5 & 51.6 &  \\
    IRCoT (Qwen3.5-27B) & 78.9 & 88.7 & 76.7 & 51.2 &  \\
    \midrule
    hypergraph walk, no reading & 74.7 & 84.7 & 71.8 & 47.8 &  \\
    \oursrow
    \method & 83.1 & 92.3 & 80.6 & 57.9 & $+8.4$ \\
    \multicolumn{6}{l}{\emph{steering and rewriting}} \\
    \quad rewriting only, no steering & 77.3 & 87.5 & 74.7 & 49.6 & $-5.7$ \\
    \quad steering only, no rewriting & 81.7 & 91.1 & 79.2 & 56.3 & $-1.3$ \\
    \quad restart not steered ($\lambda=0$) & 80.8 & 90.1 & 78.0 & 55.9 & $-2.3$ \\
    \quad transitions not steered ($\beta=0$) & 80.6 & 91.0 & 77.6 & 53.0 & $-2.4$ \\
    \multicolumn{6}{l}{\emph{a generated bridge name instead of the pointer scores}} \\
    \quad rewriting with the name & 78.1 & 88.3 & 75.3 & 50.4 & $-5.0$ \\
    \quad steering and rewriting with the name & 80.9 & 91.2 & 77.6 & 53.5 & $-2.2$ \\
    \multicolumn{6}{l}{\emph{reading}} \\
    \quad last question token only & 80.8 & 91.7 & 77.1 & 53.0 & $-2.2$ \\
    \quad re-ranking the walk instead of steering it & 80.0 & 89.8 & 78.3 & 51.4 & $-3.1$ \\
    \quad embeddings instead of attention & 76.5 & 86.1 & 73.9 & 50.2 & $-6.6$ \\
    \quad heads selected by QRHead & 77.1 & 88.0 & 73.0 & 50.0 & $-6.0$ \\
    \quad heads selected by LOCOS & 75.9 & 86.9 & 72.1 & 47.8 & $-7.2$ \\
    \quad three rounds instead of two & 83.8 & 92.5 & 81.1 & 60.8 & $+0.8$ \\
    \multicolumn{6}{l}{\emph{index}} \\
    \quad labels not in the restart & 79.9 & 91.0 & 77.3 & 48.8 & $-3.2$ \\
    \quad labels replaced by entity names & 80.9 & 91.2 & 78.0 & 53.0 & $-2.2$ \\
    \quad labels appended to the passages & 82.8 & 92.4 & 80.7 & 55.5 & $-0.3$ \\
    \bottomrule
  \end{tabular}
\end{table}

\begin{table}[t]
  \centering
  \caption{Cost on MuSiQue. Index: generated tokens (millions) and GPU hours for 11,656 passages, including head selection for \method. Per
  question: mean over the first 100 test questions. IRCoT retrieves from the dense index (--).}
  \label{tab:cost}
  \footnotesize
  \setlength{\tabcolsep}{3.5pt}
  \begin{tabular}{lccccc}
    \toprule
    & & \multicolumn{2}{c}{index} & \multicolumn{2}{c}{per question} \\
    \cmidrule(lr){3-4}\cmidrule(lr){5-6}
    Method & R@5 & M tokens & hours & tokens & seconds \\
    \midrule
    NV-Embed-v2 & 69.9 & 0 & 0.07 & 0 & 0.07 \\
    RAPTOR & 59.4 & 1.5 & 3.1 & 0 & 0.47 \\
    HippoRAG~2 & 75.3 & 6.1 & 4.2 & 61 & 2.80 \\
    PropRAG & 77.8 & 7.0 & 4.5 & 0 & 1.20 \\
    HGRAG & 72.9 & 4.3 & 1.3 & 34 & 1.5 \\
    LinearRAG & 73.8 & 0 & 1.2 & 0 & 0.33 \\
    IRCoT (4B) & 76.4 & -- & -- & 301 & 7.90 \\
    IRCoT (27B) & 78.9 & -- & -- & 482 & 11.8 \\
    \midruleours
    \oursrowlast
    \method & \textbf{83.1} & 2.9 & 2.3 & 0 & 0.51 \\
    \bottomruleours
  \end{tabular}
\end{table}

\subsection{Decomposition and Cost}
\label{app:tables}

Table~\ref{tab:gain} lists the variants of Figure~\ref{fig:hops} and Section~\ref{sec:gain} with their Recall@5, overall and by
number of hops. Table~\ref{tab:cost} lists the costs of Figure~\ref{fig:cost}, including the tokens generated to build each index and
to answer each question.

\subsection{Results on All 1,000 Questions}
\label{app:all1000}

Table~\ref{tab:all1000} scores the methods that we run on all 1,000 questions of each benchmark, development questions included,
next to the numbers that their papers publish under the same protocol. The published numbers use NV-Embed-v2 and LLMs of 70B
parameters; our runs use Qwen3.5-27B. Our reproductions match the published numbers: NV-Embed-v2 within 0.4 points of Recall@5, and
HippoRAG~2 within 0.6 points on MuSiQue and HotpotQA and 3.0 points above on 2Wiki.

\begin{table}[h]
  \centering
  \caption{Recall@5 (\%) on all 1,000 questions: our runs and published numbers under the same protocol.}
  \label{tab:all1000}
  \footnotesize
  \setlength{\tabcolsep}{2.6pt}
  \begin{tabular}{lcccccc}
    \toprule
    & \multicolumn{3}{c}{our run} & \multicolumn{3}{c}{published} \\
    \cmidrule(lr){2-4}\cmidrule(lr){5-7}
    Method & MuSiQue & 2Wiki & Hotpot & MuSiQue & 2Wiki & Hotpot \\
    \midrule
    NV-Embed-v2 & 69.3 & 76.7 & 94.4 & 69.7 & 76.5 & 94.5 \\
    RAPTOR & -- & -- & -- & 57.8 & 66.2 & 86.9 \\
    HippoRAG~2 & 74.5 & 93.4 & 95.7 & 74.7 & 90.4 & 96.3 \\
    HGRAG & -- & -- & -- & 74.1 & 93.0 & 95.5 \\
    PropRAG & -- & -- & -- & 78.3 & 94.1 & 97.4 \\
    IRCoT (4B) & 75.9 & 90.9 & 93.8 & -- & -- & -- \\
    IRCoT (27B) & 78.4 & 92.6 & 94.7 & -- & -- & -- \\
    \midruleours
    \oursrowlast
    \method & 82.9 & 97.6 & 98.4 & -- & -- & -- \\
    \bottomruleours
  \end{tabular}
\end{table}

\subsection{Single-Hop Questions}
\label{app:singlehop}

Following HippoRAG~2, Table~\ref{tab:singlehop} adds the single-hop questions of NQ~\cite{nq} and PopQA~\cite{popqa}, all 1,000 of
each. Our run of NV-Embed-v2 is within 0.4 points of the published 75.4 and 51.0. \method ranks first on both datasets, 2.7 points of
Recall@5 above HippoRAG~2 on NQ and 12.5 above PropRAG on PopQA. On PopQA, 38.1\% of the questions have complete evidence in the top
five, against 3.0\% with dense retrieval.

\begin{table}[h]
  \centering
  \caption{Recall@5 and All@5 (\%) on all 1,000 single-hop questions of NQ and of PopQA. Published: reported by HippoRAG~2 and PropRAG with
  Llama-3.3-70B (--: not reported).}
  \label{tab:singlehop}
  \footnotesize
  \setlength{\tabcolsep}{4pt}
  \begin{tabular}{lcccc}
    \toprule
    & \multicolumn{2}{c}{NQ} & \multicolumn{2}{c}{PopQA} \\
    \cmidrule(lr){2-3}\cmidrule(lr){4-5}
    Method & R@5 & All@5 & R@5 & All@5 \\
    \midrule
    \multicolumn{5}{l}{\emph{Published}} \\
    RAPTOR & 68.3 & -- & 48.7 & -- \\
    HippoRAG~2 & 78.0 & -- & 51.7 & -- \\
    PropRAG & 77.9 & -- & 56.2 & -- \\
    \multicolumn{5}{l}{\emph{Our runs}} \\
    NV-Embed-v2 & 75.1 & 42.5 & 51.4 & 3.0 \\
    hypergraph walk, no reading & 79.4 & 49.1 & 59.2 & 19.0 \\
    \midruleours
    \oursrowlast
    \method & \textbf{80.7} & \textbf{50.4} & \textbf{68.7} & \textbf{38.1} \\
    \bottomruleours
  \end{tabular}
\end{table}

\subsection{Baselines}
\label{app:baselines}

All baselines run on the corpora and questions of HippoRAG~2. LLMs are served with vLLM, decode greedily, and run without thinking
mode; the reader of \method runs in Transformers. HippoRAG~2 uses the configuration of its released script, with Qwen3.5-27B for open
information extraction and for the filter of the seed facts. Its NV-Embed-v2 is loaded in half precision, as in the encoder of
\method. IRCoT runs the loop and prompts released with HippoRAG over its dense NV-Embed-v2 retriever. As in the original IRCoT, it
keeps only the first sentence of each generated thought. RAPTOR, PropRAG, and LinearRAG run with their official code, and HGRAG,
which releases no code, is reimplemented following its paper. All four use NV-Embed-v2 and, wherever they need an LLM, Qwen3.5-27B.
ICR and QRHead run their released scoring code with Qwen3-4B on the top twenty passages of NV-Embed-v2. ICR uses its default
calibrated scoring and places the most relevant passage last. QRHead uses sixteen heads that its released detection procedure selects
for Qwen3-4B on its NQ data. Both rebuild the attention of the question tokens from their query states and the cached keys, as their
default code does.

\end{document}